\documentclass[12pt,oneside]{article}
\usepackage{epsfig, cite}
\usepackage{color}
\usepackage{ifthen}
\usepackage{graphicx}
\usepackage{amsmath}
\usepackage{txfonts}

\def\half{{1\over 2}}
\def\({\left(}
\def\){\right)}
\def\[{\left[}
\def\]{\right]}

\def\e{\begin{equation}}
\def\q{\end{equation}}
\def\m{\begin{eqnarray}}
\def\n{\end{eqnarray}}

\begin{document}
\thispagestyle{empty} \setcounter{page}{0}
\renewcommand{\theequation}{\thesection.\arabic{equation}}

\begin{flushright}

%\hfill{NEU-COSMO-09012}\\

\end{flushright}
%\vspace{4mm}
\vspace{1cm}

\begin{center}
{\huge Theoretical Limits on Agegraphic Quintessence from Weak
Gravity Conjecture}

\vspace{1.4cm}

Xiang-Lai Liu, Jingfei Zhang and Xin Zhang

\vspace{.2cm}

{\em Department of Physics, College of Sciences, Northeastern
University, Shenyang 110004, China} \\

\vspace{.2cm}

%\centerline{{\tt lai_0918@126.com}}

%\centerline{{\tt zhangxin@mail.neu.edu.cn}}

\end{center}

\vspace{0.5cm}

\centerline{ABSTRACT}
\begin{quote}%\small
%\vspace{.3cm}
In this paper, we investigate the possible theoretical constraint on
the parameter $n$ of the agegraphic quintessence model by
considering the requirement of the weak gravity conjecture that the
variation of the quintessence scalar field $\phi$ should be less
than the Planck mass $M_{\rm{p}}$. We obtain the theoretical upper
bound $n\lesssim 2.5$ that is inconsistent with the current
observational constraint result $2.637<n<2.983$ ($95.4\%$ CL). The
possible implications of the tension between observational and
theoretical constraint results are discussed.
\end{quote}
\baselineskip18pt

%\addtocounter{section}{1}
\noindent

\vspace{5mm}

\newpage

\setcounter{equation}{0}

%============================= section 1 ===============================================================================
\section{Introduction}\label{sec:1}

In 1998, two independent supernovae (SN) observation groups found
that our universe is undergoing an accelerated expansion at the
present stage, through the observations of distant type Ia
supernovae~\cite{SN}. This implies that there exists a mysterious
component, dark energy, which has large enough negative pressure,
responsible for the cosmic acceleration. Many other astronomical
observations, such as surveys of the large scale structure
(LSS)~\cite{LSS} and measurements of the cosmic microwave background
(CMB) anisotropy~\cite{CMB}, also firmly indicate that dark energy
is the dominant component in the present-day universe. It is
commonly believed that exploring the nature of dark energy is one of
the focuses in the realm of both cosmology and theoretical physics
today.

The most obvious candidate for dark energy is the famous Einstein's
cosmological constant $\lambda$ which has the equation of state
$w=-1$. However, as is well known, the cosmological constant is
plagued with the ``fine-tuning'' and ``cosmic coincidence'' problems
\cite{problems}. Another promising candidate for dark energy is the
dynamical scalar field, a slowly varying, spatially homogeneous
component. An example of scalar-field dark energy is the so-called
quintessence\cite{quintessence}, a scalar field $\phi$ slowly
evolving down its potential $V(\phi)$. Provided that the evolution
of the field is slow enough, the kinetic energy density is less than
the potential energy density, giving rise to the negative pressure
responsible to the cosmic acceleration. So far, in order to
alleviate the cosmological-constant problems and explain the
accelerated expansion, a wide variety of scalar-field dark energy
models have been proposed. Besides quintessence, these also include
phantom, $k$-essence, tachyon, ghost condensate and quintom amongst
many. However, we should note that the mainstream viewpoint regards
the scalar-field dark energy models as a low-energy effective
description of the underlying theory of dark energy.

It is generally believed by theorists that we cannot entirely
understand the nature of dark energy before a complete theory of
quantum gravity is established. However, although we are lacking a
quantum gravity theory today, we still can make some efforts to
explore the nature of dark energy according to some principles of
quantum gravity. The holographic dark energy model \cite{Li:2004rb}
is just an appropriate example, which is constructed in light of the
holographic principle of quantum gravity theory. That is to say, the
holographic dark energy model possesses some significant features of
an underlying theory of dark energy. More recently, a new model
consistent with the holographic principle, the agegraphic dark
energy model, has been proposed in ~\cite{Cai:2007us}, which takes
into account the uncertainty relation of quantum mechanics together
with the gravitational effect in general relativity.

While, by far, a complete theory of dark energy has not been
established presently, we can, however, speculate on the underlying
theory of dark energy by taking some principles of quantum gravity
into account. The agegraphic dark energy model is no doubt a
tentative in this way. Now, we are interested in that if we assume
the holographic/agegraphic vacuum energy scenario as the underlying
theory of dark energy, how the low-energy effective scalar-field
model can be used to describe it. In this direction, some work has
been done, see, e.g.,
\cite{holoscalar,Zhang:2009un,Zhang:2008mb,Cui:2009ns}. The
agegraphic versions of scalar-field models, such as quintessence and
tachyon, have been constructed \cite{Zhang:2008mb,Cui:2009ns}. In
this paper, we focus on the canonical scalar-field description of
the agegraphic dark energy, namely, the ``agegraphic quintessence''
\cite{Zhang:2008mb}.

In recent years, cosmological-constant/dark-energy problem has been
studied by string theorists within the string framework. It is
generally considered that string theory is the most promising
consistent theory of quantum gravity. Based on the KKLT mechanism
\cite{Kachru:2003aw}, a vast number of metastable de Sitter vacua
have been constructed through the flux compactification on a
Calabi-Yau manifold. These string vacua can be described by the
low-energy effective theories. Furthermore, it is realized that the
vast series of semiclassically consistent field theories are
actually inconsistent. These inconsistent effective field theories
are believed to locate in the so-called ``swampland''
\cite{Vafa:2005ui}. The self-consistent landscape is surrounded by
the swampland. Vafa has proposed some criterion to the consistent
effective field theories \cite{Vafa:2005ui}. Moreover, it was
conjectured by Arkani-Hamed et al. \cite{weakgrav} that the gravity
is the weakest force, which helps rule out those effective field
theories in the swampland. Arkani-Hamed et al. pointed out
\cite{weakgrav} that when considering the quantum gravity, the
gravity and other gauge forces should not be treated separately. For
example, in four dimensions a new intrinsic UV cutoff for the U(1)
gauge theory with single scalar field, $\Lambda=g M_{\rm{p}}$, is
suggested, where $g$ is the gauge coupling \cite{weakgrav}. In
\cite{hls}, the weak gravity conjecture together with the
requirement that the IR cutoff should be smaller than the UV cutoff
leads to an upper bound for the cosmological constant. In addition,
for the inflationary cosmology, the application of the weak gravity
conjecture shows that the chaotic inflation model is in the
swampland \cite{Huangchaotic}. This conjecture even implies that the
eternal inflation may not be achieved \cite{HLW}. Furthermore, Huang
conjectured \cite{Huang:2007qz} that the variation of the inflaton
should be smaller than the Planck scale $M_{\rm{p}}$, and this can
make stringent constraint on the spectral index.

More recently, the weak gravity conjecture has been applied to the
dark-energy problem. It is suggested that the variation of the
quintessence field value $\phi $ should be less than $M_{\rm{p}}$.
This criterion may give important theoretical constraints on the
equation-of-state parameter of quintessence models, and some of
these constraints are even stringent than those of the present
experiments \cite{Huang:2007mv}. The criterion
$|\Delta\phi(z)|/M_{\rm{p}}\leq 1$ has also been used to put
theoretical constraints on other canonical scalar-field dark energy
models; see, e.g., \cite{Ma:2007av,Wu:2007pq,Chen:2008gi}. In this
paper we shall investigate the possible theoretical limits on the
parameter $n$ of the agegraphic quintessence from the weak gravity
conjecture.

In the next section, we will briefly review the new agegraphic dark
energy model proposed in~\cite{Wei:2007ty}. In Sec.~\ref{sec3}, we
will give the possible theoretical limits on the parameter $n$ of
the agegraphic quintessence model from the weak gravity conjecture.
Conclusion will be given in Sec.~\ref{sec4}.

%============================= section 2 =================================================================================
\setcounter{equation}{0}
\section{Agegraphic Dark Energy Model}\label{sec2}

Holographic dark energy models arise from the holographic principle
\cite{Hooft93} of quantum gravity. The holographic principle
determines the range of validity for a local effective quantum field
theory to be an accurate description of the world involving dark
energy, by imposing a relationship between the ultraviolet (UV) and
infrared (IR) cutoffs \cite{Cohen:1998zx}. As a consequence, the
vacuum energy becomes dynamical, and its density $\rho_{\rm de}$ is
inversely proportional to the square of the IR cutoff length scale
$L$ that is believed to be some horizon size of the universe,
namely, $\rho_{\rm de}\propto L^{-2}$.

The original holographic dark energy model \cite{Li:2004rb} chooses
the future event horizon size as its IR cutoff scale, so the energy
density of holographic dark energy reads $\rho_{\rm de}=3c^2M^2_{\rm
p}R_{\rm eh}^{-2}$, where $c$ is a constant, and $R_{\rm eh}$ is the
size of the future event horizon of the universe. This model is
successful in explaining the cosmic acceleration and in fitting the
observational data. There are also other two versions of holographic
dark energy, namely, the agegraphic dark energy model
\cite{Cai:2007us,Zhang:2008mb,Cui:2009ns,Wei:2007ty,nadeext} and the
holographic Ricci dark energy model \cite{Zhang:2009un,rde}. In this
paper, we focus on the agegraphic dark energy model.

The agegraphic dark energy model discussed in this paper is actually
the new version of the agegraphic dark energy model
\cite{Wei:2007ty} (sometimes called the new agegraphic dark energy
model in the literature) which suggests to choose the conformal age
of the universe
\begin{equation}
\eta=\int_0^t \frac{dt'}{a}=\int_0^a \frac{da'}{Ha'^{2}}
\label{eq13}
\end{equation}
as the IR cutoff, so the energy density of agegraphic dark energy is
\begin{equation}
\rho_{\rm de}=3n^{2}M_{\rm p}^{2}\eta^{-2}, \label{eq12}
\end{equation}
where $n$ is a constant which plays the same role as $c$ in the
original holographic dark energy model.

The corresponding fractional energy density is given by
\begin{equation}
\Omega_{\rm{de}}=\frac{n^2}{H^2\eta^2}. \label{eqO}
\end{equation}
Taking derivative for Eq. (\ref{eqO}) with respect to $x=\ln a$, and
considering Eq.~(\ref{eq13}), we obtain
\begin{equation}
\Omega'_{\rm{de}}=2\Omega_{\rm{de}}\left(\epsilon-{\sqrt{\Omega_{\rm{de}}}\over
na}\right).
\end{equation}
where $\epsilon\equiv -{\dot{H}/ H^2}$. The Friedmann equation reads
\begin{equation}
3M_{\rm{p}}^2H^2=\rho_{\rm{m}}+\rho_{\rm{de}},\label{Fri}
\end{equation}
or equivalently,
\begin{equation}
E(z)\equiv {H(z)\over H_0}=\left(\Omega_{\rm{m0}}(1+z)^3\over
1-\Omega_{\rm{de}}\right)^{1/2}.\label{Ez}
\end{equation}
From Eqs.~(\ref{Fri}), (\ref{eq12}), (\ref{eqO}) and
$\dot{\rho}_{m}+3H\rho_{m}=0$, we have
\begin{equation}
\epsilon={3\over
2}({1-\Omega_{\rm{de}}})+{\Omega_{\rm{de}}^{3/2}\over na}.
\end{equation}
Hence, we get the equation of motion for $\Omega_{\rm{de}}$, i.e.,
\begin{equation}
\Omega
_{\rm{de}}^{\prime}=\Omega_{\rm{de}}(1-\Omega_{\rm{de}})\left(3-\frac{2}{n}\frac{\sqrt{\Omega_{\rm{de}}}}{a}\right)~,
\end{equation}
and this equation can be rewritten as
\begin{equation}
\frac{d
\Omega_{\rm{de}}}{dz}=-\Omega_{\rm{de}}(1-\Omega_{\rm{de}})\left(3(1+z)^{-1}-\frac{2}{n}\sqrt{\Omega_{\rm{de}}}\right)~\label{keyq}.
\end{equation}
From Eqs. (\ref{eq12}), (\ref{eqO}) and
$\dot{\rho}_{\rm{de}}+3H(1+w_{\rm{de}})\rho_{\rm{de}}=0$, we obtain
the equation of state (EoS) of the agegraphic dark energy
\begin{equation}
w_{\rm{de}}=-1+{2\over 3n}{\sqrt{\Omega_{\rm{de}}}\over
a}.\label{wade}
\end{equation}

Now, we pause for a while to make some additional comments on the
old version of the agegraphic dark energy model \cite{Cai:2007us}.
In the old model, the IR cutoff of the theory is taken as the age of
the universe, $t=\int_0^a {da\over Ha}$. However, for this choice,
there are some internal inconsistencies in the model; see
\cite{Wei:2007ty} for detailed discussions. In the matter-dominated
epoch with $\Omega_{\rm{de}}\ll 1$, one has $a\propto t^{2/3}$, thus
$t^2\propto a^3$. So, in this epoch, $\rho_{\rm{de}}\propto
t^{-2}\propto a^{-3}$. Since $\rho_{\rm{m}}\propto a^{-3}$, one has
$\Omega_{\rm{de}}\simeq {\rm const.}$, which is conflict with
$\Omega_{\rm{de}}\propto a^3$ obtained from the differential
equation governing the evolution of dark energy \cite{Wei:2007ty}.
What's more, from $\rho_{\rm{de}}\propto t^{-2}$, the agegraphic
dark energy tracks the dominated components (either pressureless
matter or radiation). Therefore, the agegraphic dark energy never
dominates. This is of course unacceptable. Accordingly, the new
version of the agegraphic dark energy model was proposed
\cite{Wei:2007ty} by replacing the age $t$ with the conformal age
$\eta$, for eliminating the inconsistencies in the old version. This
is the reason why we only consider the new agegraphic dark energy
model in this paper.

%=================================section 3.0=============================================================================
\setcounter{equation}{0}
\section{Agegraphic Quintessence and Its Possible Theoretical Limits from Weak Gravity
Conjecture}\label{sec3}

For a single-scalar-field quintessence model, the potential energy
density $V(\phi)$ is a function of the scalar field $\phi$. If the
field is spatially homogeneous, namely, the spacial curvature of
field can be neglected, the field equation can be expressed as
\begin{equation}
\ddot{\phi} + 3H\dot\phi + {dV\over d\phi } = 0, \label{eq:fieldeq}
\end{equation}
where the dot denotes the derivative with respect to the cosmic
time. The energy density and the pressure are
\begin{eqnarray}
\rho_\phi & = & {1\over 2} \dot\phi^2 + V(\phi), \nonumber \\
p_\phi & = & {1\over 2} \dot\phi^2 - V(\phi), \label{eq:rhophi}
\end{eqnarray}
so the EoS parameter is
\begin{equation}
w_{\phi}={p_\phi\over \rho_\phi}={\half {\dot \phi}^2-V(\phi)\over
\half {\dot \phi}^2+V(\phi)}, \label{eosp}
\end{equation}
which generally varies with time. The range for the EoS parameter of
the quintessence is $w_{\phi}\in [-1,1]$. If the scalar field varies
slowly in time, namely, $\dot\phi^2 \ll V$, the field energy
approximates the effect of Einstein's cosmological constant with
$p_\phi\simeq - \rho_\phi$.

Using Eq. (\ref{eosp}), we find a relationship between the potential
of quintessence and its kinetic energy
\begin{equation}
V(\phi)={{\dot\phi}^2\over 2}{1-w_{\phi}\over 1+w_{\phi}}.
\end{equation}
The energy density takes the form
\begin{equation}
\rho_\phi=\half{\dot \phi}^2+V(\phi)={{\dot \phi}^2\over
1+w_{\phi}}. \label{rw}
\end{equation}
We assume, without loss of generality, $dV/d\phi>0$, so that
$\dot\phi<0$. Thus, Eq. (\ref{rw}) reads
\begin{equation}
\dot \phi=-\sqrt{(1+w_{\phi})\rho_\phi}.\label{dp}
\end{equation}
In the agegraphic quintessence model \cite{Zhang:2008mb}, the
quintessence scalar field is viewed as an effective description of
the agegraphic dark energy, so the scalar-field energy density
$\rho_\phi$ and EoS $w_\phi$ are identified with those of the
agegraphic dark energy, $\rho_{\rm{de}}$ and $w_{\rm{de}}$,
respectively.

Integrating Eq. (\ref{dp}), we obtain
\begin{eqnarray}
{|\Delta \phi(z)|\over M_{\rm{p}}}&=&\int_{\phi(0)}^{\phi(z)} d\phi/M_{\rm{p}}\nonumber \\
&=&\int_0^{z}\sqrt{3[1+w_{\rm{de}}(z')]\Omega_{\rm{de}}(z')}{dz'\over
1+z'}, \label{vp1}
\end{eqnarray}
where $\Omega_{\rm{de}}$ and $w_{\rm{de}}$ are given by
Eqs.~(\ref{keyq}) and (\ref{wade}) for the agegraphic quintessence
model. If we fix the field amplitude at the present epoch ($z=0$) to
be zero, $\phi(0)=0$, then Eq. (\ref{vp1}) can be rewritten as
\begin{equation}
{\phi(z)\over
M_{\rm{p}}}=\int_0^{z}\sqrt{3[1+w_{\rm{de}}(z')]\Omega_{\rm{de}}(z')}{dz'\over
1+z'}.\label{vph1}
\end{equation}

As suggested in \cite{Caldwell:2005tm}, quintessence models can be
divided into two classes, ``thawing'' models and ``freezing''
models. Thawing models depict those scalar fields that evolve from
$w=-1$ but grow less negative with time as $dw/d\ln a>0$; freezing
models, whereas, describe those fields that evolve from $w>-1$ and
$dw/d\ln a<0$ to $w\rightarrow -1$ and $dw/d\ln a\rightarrow 0$. The
agegraphic dark energy mimics a cosmological constant at the late
time, so it belongs to the freezing quintessence models
\cite{Zhang:2008mb}. A particular feature of this model is that it
is actually a single-parameter model: the differential equation of
$\Omega_{\rm{de}}$, namely, Eq. (\ref{keyq}), is governed by a
single parameter $n$, provided that the initial condition is taken
to be $\Omega_{\rm{de}}(z_{\rm ini})=n^2(1+z_{\rm ini})^{-2}/4$ at
any $z_{\rm ini}$ which is deep enough into the matter-dominated
epoch. Following \cite{Wei:2007ty}, here we take $z_{\rm ini}=2000$.

%===================================Fig.1================================================================================
\begin{figure}[tbph]
\begin{center}
\includegraphics[width=0.60\textwidth]{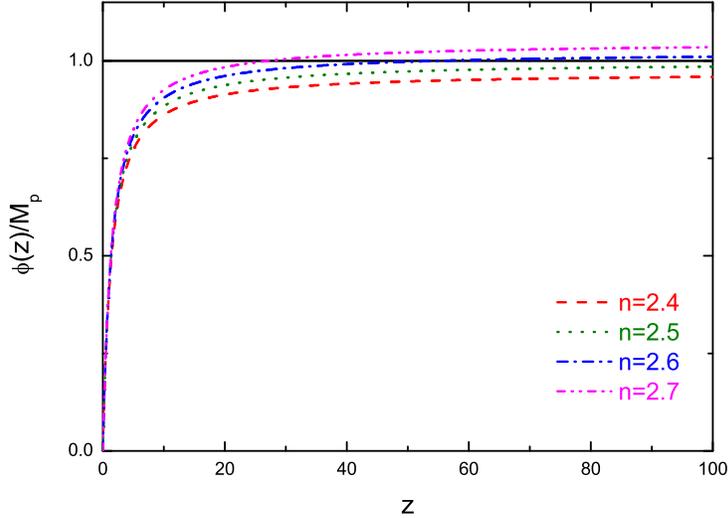}
\end{center}
\caption{\label{fig1} \small The scalar field evolution for the
single-field agegraphic quintessence model. The theoretical
requirement $|\Delta\phi(z)|/M_{\rm{p}}\leq 1$ places a constraint
on this model, $n\lesssim 2.5$, which is inconsistent with the
current observational constraint $2.637<n<2.983$.} \label{fig1}
\end{figure}

It should be mentioned that the agegraphic dark energy model has
been constrained strictly by using the latest observational data
including the Constitution sample of SN, the shift parameter of the
CMB given by the five-year Wilkinson Microwave Anisotropy Probe
(WMAP) observations, and the baryon acoustic oscillation (BAO)
measurement from the Sloan Digital Sky Survey (SDSS)
\cite{Li:2009bn}. The analysis of these observational data gives the
fitting results \cite{Li:2009bn}: for $68.3\%$ confidence level,
$n=2.807^{+0.087}_{-0.086}$; for $95.4\%$ confidence level,
$n=2.807^{+0.176}_{-0.170}$.

Consider the theoretical constraint on the single-field agegraphic
quintessence model in which the variation of the canonical scalar
field $|\Delta\phi(z)|$ is required not to exceed the Planck scale
$M_{\rm{p}}$. Figure~\ref{fig1} shows the constraint result:
$n\lesssim 2.5$, which is a surprising result because this limit is
so stringent and somewhat inconsistent with the result obtained from
the current observational data. According to the current
observations, at $95.4\%$ confidence level, we have $2.637<n<2.983$
\cite{Li:2009bn} that refuses to accommodate the theoretical limit
$n\lesssim 2.5$.

One may naturally ask whether the multi-field agegraphic
quintessence model could loosen the theoretical limit and eliminate
the above tension between theoretical and observational limits. In
the following, we shall give a clear answer to this question.

Let us consider a quintessence scalar-field model containing $N$
scalar fields $\phi_i$ with independent potential $V_i(\phi_i)$ for
$i=1,~\dots,~N$. Thus, for each scalar field $\phi_i$, we have
\begin{equation}
\ddot{\phi_{i}} + 3{H}\dot\phi_{i} + {dV_{i}(\phi_{i})\over
d\phi_{i}} = 0, \label{eq:fieldeq}
\end{equation}
where the dot denotes the derivative with respect to the cosmic
time. The total energy density and the pressure of the fields are
\begin{eqnarray}
\rho_\phi& = & {1\over 2}\sum_{i=1}^{N}\dot\phi_{i}^2 +
\sum_{i=1}^{N}V_{i}(\phi_{i}), \nonumber
\\ p_\phi & = & {1\over 2}\sum_{i=1}^{N}\dot\phi_{i}^2 -
\sum_{i=1}^{N}V_{i}(\phi_{i}). \label{eq1:rhophi}
\end{eqnarray}
For simplicity, we assume that $\phi_{1}=\phi_{2}=\ldots
=\phi_{i}=\ldots=\phi_{N}\equiv\varphi$ and $V_{1}(\phi_{1})=
V_{2}(\phi_{2})=\ldots=V_{i}(\phi_{i})=\ldots=V_{N}(\phi_{N})\equiv
V(\varphi)$. Then, the total energy density and the pressure of the
scalar fields can be rewritten as
\begin{eqnarray}
\rho_\phi& = & N\left({1\over 2}\dot\varphi^2 + V(\varphi)\right),
\nonumber
\\ p_\phi& = & N\left({1\over 2}\dot\varphi^2 -
V(\varphi)\right), \label{eq2:rhophi}
\end{eqnarray}
and the EoS parameter can be expressed as
\begin{equation}
w_\phi={p_\phi\over \rho_\phi}=\frac{{1\over 2}\dot\varphi^2 -
V(\varphi)}{{1\over 2}\dot\varphi^2 + V(\varphi)}. \label{eos}
\end{equation}
Using Eqs.~(\ref{eq2:rhophi}) and (\ref{eos}), we obtain
\begin{equation}
\dot \varphi=-\sqrt{\frac{1}{N}(1+w_\phi)\rho_\phi}.\label{dpw}
\end{equation}
Note that in this expression we have assumed $dV/d\varphi>0$, in
accordance with the previous discussion. Next, we identify the
scalar-field energy density $\rho_\phi$ and EoS $w_\phi$ with those
of the agegraphic dark energy, $\rho_{\rm{de}}$ and $w_{\rm{de}}$,
respectively. Integrating Eq. (\ref{dpw}), we obtain
\begin{eqnarray}
{|\Delta \varphi(z)|\over M_{\rm p}}&=&\int_{\varphi(0)}^{\varphi(z)} d\varphi/M_{\rm{p}}\nonumber \\
&=&\int_0^{z}\sqrt{\frac{3[1+w_{\rm{de}}(z')]\Omega_{\rm{de}}(z')}{N}}{dz'\over
1+z'}. \label{vph}
\end{eqnarray}
It is easy to see that in this case the amplitude of $|\Delta
\varphi(z)|$ is suppressed by a factor $N^{-1/2}$, so it seems that
the tension between the theoretical and observational limits in the
single-field case could be avoided in such a multi-field model.

Fixing the field amplitude at the present epoch to be zero,
$\varphi(0)=0$, from Eq.~(\ref{vph}) we get
\begin{equation}
{\varphi(z)\over M_{\rm
p}}=\int_0^{z}\sqrt{\frac{3[1+w_{\rm{de}}(z')]\Omega_{\rm{de}}(z')}{N}}{dz'\over
1+z'}.\label{vph1}
\end{equation}
Furthermore, using Eqs.~(\ref{eq2:rhophi}) and (\ref{eos}), we
obtain
\begin{eqnarray}
{V(\varphi)}& = &\frac{1}{2N}(1-w_\phi)\rho_\phi \nonumber
\\ &=& \frac{\rho_{\rm c0}}{2N}(1-w_\phi)\Omega_\phi E^{2},
\end{eqnarray}
or equivalently,
\begin{equation}
\frac{{V(\varphi)}}{\rho_{\rm c0}}=
\frac{1}{2N}(1-w_{\rm{de}})\Omega_{\rm{de}}E^{2},\label{vp}
\end{equation}
where $\rho_{\rm c0}=3M_{p}^{2}H_{0}^{2}$ is today's critical
density of the universe.

\begin{center}
\begin{figure}[htbp]
%\centering
\includegraphics[width=0.5\textwidth]{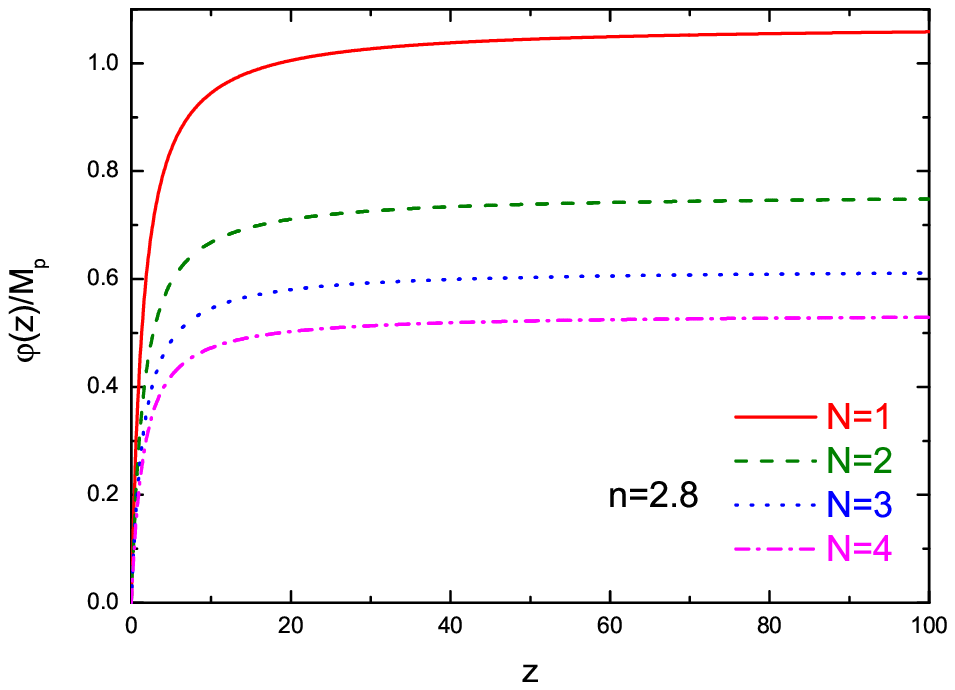}\hfill
\includegraphics[width=0.5\textwidth]{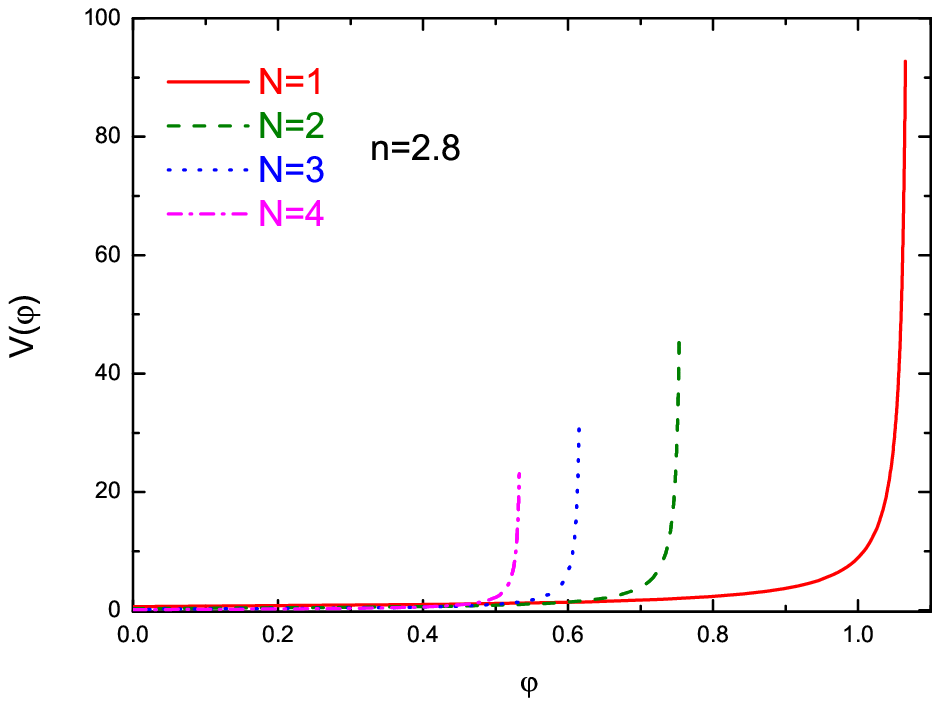}
\caption{\label{fig2} \small Multi-field agegraphic quintessence
model corresponding to $n=2.8$. In the left panel, we plot the
evolution of the scalar field $\varphi(z)$; in the right panel, we
show the corresponding potential $V(\varphi)$. Note that in the
right panel the potential $V(\varphi)$ is in unit of $\rho_{\rm c0}$
and the field $\varphi$ is in unit of $M_{\rm p}$. It is clear to
see from this figure that bigger $N$ indeed gives rise to smaller
$|\Delta\varphi|$.}
\end{figure}
\end{center}

\begin{figure}[tbph]
\begin{center}
\includegraphics[width=0.5\textwidth]{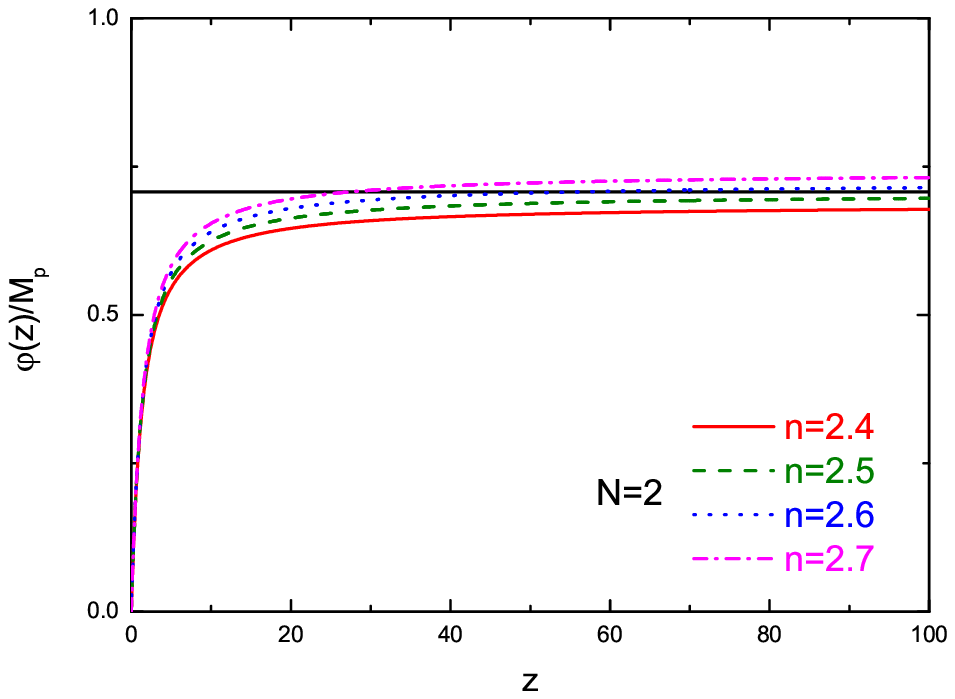}\hfill
\includegraphics[width=0.5\textwidth]{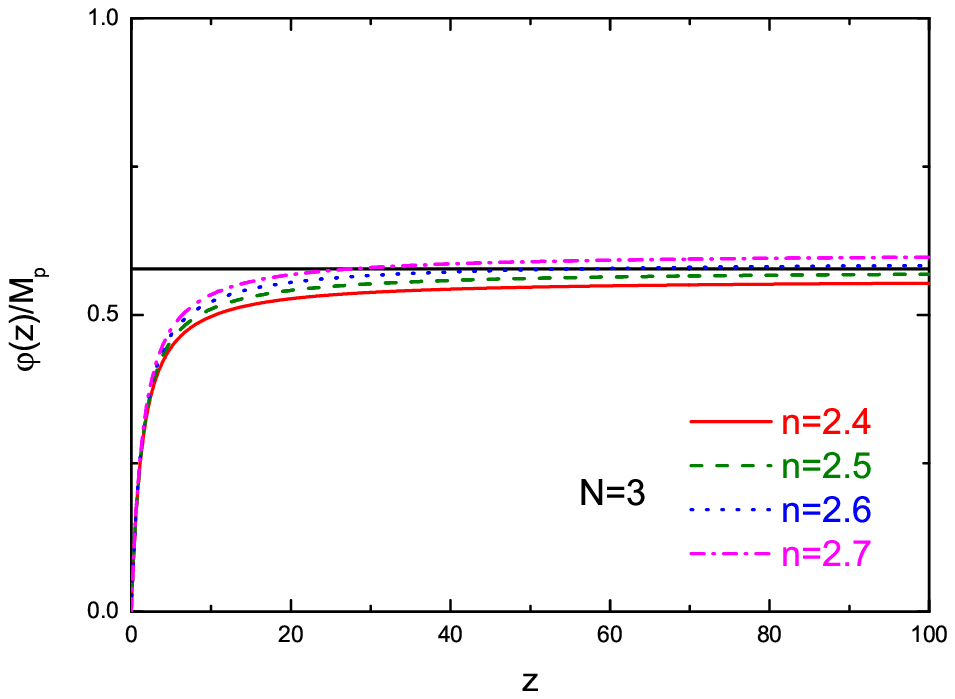}
\end{center}
\caption{\label{fig3} \small The scalar field evolution $\varphi(z)$
for the multi-field agegraphic quintessence model. Here, we show two
concrete examples, $N=2$ and 3. The weak gravity conjecture for
multi-field theory indicates that the same theoretical limit
$n\lesssim 2.5$, independent of $N$, will be given.}
\end{figure}

By far, we have constructed a multi-field quintessence model
mimicking the agegraphic dark energy. Figure \ref{fig2} is an
example of multi-field agegraphic quintessence model corresponding
to $n=2.8$. In the left panel, we plot the evolution of the scalar
field $\varphi(z)$; the corresponding potential $V(\varphi)$ can be
found in the right panel. From this figure, it is clear to see that
bigger $N$ indeed gives rise to smaller $|\Delta\varphi|$. So, if we
use $|\Delta\varphi(z)|/M_{\rm{p}}\leq 1$ to constrain the
multi-field agegraphic quintessence model, the tension between the
theoretical and observational limits in the single-field case would
be removed. However, does the criterion
$|\Delta\varphi(z)|/M_{\rm{p}}\leq 1$ still hold in a multi-field
model? Unfortunately, the answer is NO. It has been demonstrated by
Huang \cite{weakgrav2} that for an effective canonical scalar field
theory with $N$ species the weak gravity conjecture requires that
the maximal variation of the scalar field satisfies the bound
$|\Delta\varphi|/M_{\rm p}\lesssim 1/\sqrt{N}$, i.e., the upper
bound of the field variation in the multi-field case is also
suppressed by a factor $N^{-1/2}$.\footnote{In Ref.~\cite{weakgrav2}
the author proposes the weak gravity conjecture for a multiple
scalar field theory and finds that the variation of the canonical
scalar field is bounded by the gravity scale $\Lambda_G=M_{\rm
p}/\sqrt{N}$, when an unimportant coefficient 2 is ignored. In this
paper we also ignore this unimportant coefficient in order to keep
the whole work consistent.} Therefore, obviously, the weak gravity
conjecture for the multi-field agegraphic quintessence model must
give the same theoretical limit result as the single-field model.
For clarity, we illustrate two concrete examples, $N=2$ and 3, in
Fig.~\ref{fig3}. In this figure, we explicitly show that the same
theoretical limit $n\lesssim 2.5$ is obtained for the multi-field
agegraphic quintessence model from weak gravity conjecture.

%============================= Fig.2======================================================================

%===================================Fig.3================================================================================

Therefore, the agegraphic quintessence model is facing an awkward
situation that the theoretical limit derived from weak gravity
conjecture, $n\lesssim 2.5$, is not in accordance with the
observational constraint result, $2.637<n<2.983$ (95.4\% CL). In
fact, if the weak gravity conjecture is taken seriously, some
low-energy effective field theories have been demonstrated to be in
the swampland. For example, not only the chaotic inflation model is
in the swampland \cite{Huangchaotic}, but also the assisted chaotic
inflation might not be in the landscape \cite{weakgrav2}. The
$N$-flation, a possible realization of the assisted chaotic
inflation in string theory, is shown to be just semiclassically
self-consistent, not really self-consistent, if the weak gravity
conjecture is correct \cite{weakgrav2}. In the present paper, we
find that the weak gravity conjecture leads to a tension between the
theoretical and observational limits for the agegraphic dark energy
model. This inconsistency can be explained as that the agegraphic
dark energy might not be described by a consistent low-energy
effective scalar field theory. Of course, one may argue that this
assertion looks too strong, since after all the theoretical and
observational limits both locate in around $n\sim 2-3$, which can
also be viewed as that they are, to some extent, not in conflict.
So, for the sake of rigorous, we do not exclude other possibilities
such as that the current observational data cannot provide precise
constraint on the agegraphic dark energy model, and the future
accurate data perhaps would change the constraint result to be
consistent with the theoretical limit.

%============================= section 4 ==================================================================================
\setcounter{equation}{0}
\section{Conclusion}\label{sec4}

To summarize, in this paper we have investigated the theoretical
limits on the parameter $n$ of the agegraphic quintessence model by
considering that the variation of the quintessence scalar field
$\phi$ should be less than the Planck mass $M_{\rm{p}}$. The
agegraphic dark energy can mimic the behavior of a quintessence
scalar-field dark energy, so the quintessence model can be used to
effectively describe the agegraphic dark energy. In this paper, we
have tested the single-field and multi-field agegraphic quintessence
models by using the weak gravity conjecture. We believe that the
low-energy effective field theory is not applicable in the
trans-Planckian field space.

We have shown that for both single-field and multi-field agegraphic
quintessence models the weak gravity conjecture leads to the same
theoretical limit, $n\lesssim 2.5$, which is inconsistent with the
current observational constraint $2.637<n<2.983$ (95.4\% CL). The
requirement that the variation of the field should be less than the
Planck scale from weak gravity conjecture may arise from the
consistent theory of quantum gravity, so in this sense the
theoretical result obtained in this paper can, to some extent, be
viewed as the prediction of quantum gravity. The tension between
theoretical and observational limits implies that the agegraphic
dark energy could not be described by a consistent low-energy
effective scalar field theory. Of course, other possible reasons for
the tension still exist, for example, perhaps the current
observational data cannot provide precise constraint on the
agegraphic dark energy model, and the future accurate data might
change the constraint result to be consistent with the theoretical
limit from weak gravity conjecture.

\section*{Acknowledgements}
We would like to thank the referee for providing us with many
helpful suggestions. This work was supported by the National Natural
Science Foundation of China under Grant Nos.~10705041 and 10975032.

%\newpage

\end{document}